\documentclass[pdflatex,sn-mathphys-num]{sn-jnl}

\usepackage{graphicx}%
\usepackage{multirow}%
\usepackage{amsmath,amssymb,amsfonts}%
\usepackage{amsthm}%
\usepackage{mathrsfs}%
\usepackage{subfig}%
\usepackage[title]{appendix}%
\usepackage{xcolor}%
\usepackage{textcomp}%
\usepackage{manyfoot}%
\usepackage{booktabs}%
\usepackage{algorithm}%
\usepackage{algorithmicx}%
\usepackage{algpseudocode}%
\usepackage{listings}%

\theoremstyle{thmstyleone}%
\theoremstyle{thmstyletwo}%

\theoremstyle{thmstylethree}%

\begin{document}

\title[Article Title]{Pre-trained Audio Transformer as a Foundational AI Tool for Gravitational Waves}


\author*[1,2]{\fnm{Chayan} \sur{Chatterjee}}\email{chayan.chatterjee@vanderbilt.edu}

\author[2]{\fnm{Abigail} \sur{Petulante}}\email{abigail.petulante@vanderbilt.edu}

\author[2]{\fnm{Yang} \sur{Hu}}\email{yang.hu.1@vanderbilt.edu}

\author[1,2]{\fnm{Roy} \sur{Lau}}\email{roy.lau@vanderbilt.edu}

\author[1]{\fnm{Suyash} \sur{Deshmukh}}\email{suyash.deshmukh@vanderbilt.edu}

\author[2]{\fnm{Haowei} \sur{Fu}}\email{haowei.fu@vanderbilt.edu}

\author[2]{\fnm{Trang} \sur{Hoang}}\email{trang.t.hoang.1@vanderbilt.edu }

\author[2]{\fnm{Stephen Chong} \sur{Zhao}}\email{chong.zhao.1@vanderbilt.edu}

\author[2]{\fnm{Jesse} \sur{Spencer-Smith}}\email{jesse.spencer-smith@vanderbilt.edu}

\author[1]{\fnm{Karan} \sur{Jani}}\email{karan.jani@vanderbilt.edu}

\affil*[1]{\orgdiv{Department of Physics and Astronomy}, \orgname{Vanderbilt University}, \orgaddress{\street{2201 West End Avenue}, \city{Nashville}, \postcode{37235}, \state{Tennessee}, \country{USA}}}

\affil[2]{\orgdiv{Data Science Institute}, \orgname{Vanderbilt University}, \orgaddress{\street{1400 18th Avenue South Building, Suite 2000}, \city{Nashville}, \postcode{37212}, \state{Tennessee}, \country{USA}}}


\abstract{As gravitational wave detectors become more advanced and sensitive, the number of signals recorded by Advanced LIGO and Virgo from merging compact objects is expected to rise dramatically. This surge in detection rates necessitates the development of adaptable, scalable, and efficient tools capable of addressing a wide range of tasks in gravitational wave astronomy. Foundation AI models present a transformative opportunity in this context by providing a unified framework that can be fine-tuned for diverse applications while leveraging the power of large-scale pre-training. In this work, we explore how pre-trained audio foundation models, specifically OpenAI's Whisper, can be adapted for gravitational wave data analysis. By fine-tuning Whisper’s encoder model -- originally trained on extensive audio data -- we achieve reliable results in detecting gravitational wave signals and classifying transient noise artifacts or `glitches'. This represents the first cross-domain application of an open‑source audio transformer to gravitational wave research, demonstrating that models trained in one domain can be repurposed for versatile and efficient data analysis amid rising detection rates.
}

\maketitle


The discovery of gravitational waves (GWs) in 2015 \cite{GW150914} by the advanced Laser Interferometer Gravitational Wave Observatory (LIGO) \citep{LIGO} detectors marked a new era in our exploration of the cosmos, opening a unique window to study some of the most energetic events in the universe. Since then more than 90 GW events have been detected from the collisions of black holes and neutron stars \citep{GWTC1, GWTC2, GWTC_2_1, GWTC-3}, which has provided unique insights into the lives of massive stars, the behavior of matter under extreme conditions, and the nature of spacetime \citep{HubbleConstant, TGR, EOS, GWTC-3}. With ongoing improvements in the sensitivity of the LIGO and Virgo \citep{Virgo} detectors and the planned deployment of next-generation observatories \citep{3G_detectors, 3G_detectors_1}, the number of detected events is projected to rise dramatically, potentially yielding several observations each day \citep{Detections_3G_era}. This rapid growth presents significant computational challenges due to the vast volume of data generated which requires sophisticated analyses to extract astrophysical information. Traditional methods for GW data analysis, such as matched filtering \citep{Matched_filtering} and Bayesian parameter estimation \citep{Bilby, LALInference}, while effective, are computationally intensive and struggle to scale with the growing detection rates. \\

Artificial intelligence (AI) has emerged as a powerful solution to the growing challenges of GW data analysis. Deep learning models \citep{Deep_learning_Goodfellow} can now match the sensitivity of traditional pipelines while reducing computational costs, and have been successfully applied to rapid signal detection \citep{detection2,ML_MDC,Gabbard_detection,aframe,NSBH_detection_ML,BNS_detection_ML,MLy_pipeline}, parameter estimation \citep{DINGO_IS,Gabbard,Nessai,Chua,ML_PE_Shen}, and the classification of non‑stationary noise transients, or ``glitches”, whose ever‑changing morphologies and detector–specific signatures pose a constant challenge to model‑based searches \citep{glitches_definition1,glitches_definition2,glitches_definition3,GravitySpy,GravitySpy_classifier}. However, because these disturbances evolve over time and can mimic astrophysical signals, truly robust, model‑agnostic methods are needed that generalize across noise conditions, detector configurations, and signal morphologies without requiring exhaustive retraining. Foundation models, such as large, open‑source pre‑trained architectures \citep{LLM1,LLM2,LLM3}, offer precisely this capability: having already learned rich, transferable representations from massive audio datasets, they can be efficiently fine‑tuned to GW-specific tasks. This allows us to leverage powerful feature extractors while dramatically reducing training overhead. \\


In this article, we introduce GW-Whisper, the first application of OpenAI’s Whisper model \citep{Whisper} for GW data analysis. Whisper is a transformer-based model \citep{Transformer} and is the current state-of-the-art foundation automatic speech recognition (ASR) system designed for robust transcription and translation across diverse languages and acoustic conditions. Trained on 680,000 hours of multilingual and multitask supervised data, Whisper excels in noisy environments and supports speech-to-text tasks, including language detection and transcription with high accuracy \citep{Whisper}. Since Whisper is trained on audio data, which predominantly lies in the same frequency range as the sensitive LIGO frequency range (10 - 10,000 Hz), we explore its application to GW data analysis. This work highlights the potential of repurposing state-of-the-art ASR systems for advancing the frontiers of GW astrophysics. In this work, we apply GW-Whisper for two tasks - GW detection and multi-class classification of glitches observed in LIGO data.\\

Whisper (from \texttt{openai/whisper-tiny}) was originally trained on 30 s human-speech clips resampled to 16 kHz, converting each waveform into an 80-band log-mel spectrogram (25 ms windows, 10 ms hops) before processing through two down-sampling convolutional layers \citep{log-mel-spectrogram,CNN_1} and a 24-layer Transformer encoder \citep{Transformer,Whisper}. To adapt to $\mathcal{O}(1\rm\,s)$ GW signals spanning $\sim20\text{–}1024$ Hz, GW-Whisper replaces the log-mel front end with a per-detector Q-transform (QScan) followed by a lightweight “Q-Adapter”: three convolutional blocks (Conv → ReLU → Pool) that reshape the spectrogram into the encoder’s expected $80\times3000$ input. The Q-Adapter output is then modulated via detector-specific feature-wise linear modulation (FiLM) parameters $\gamma_i$ and $\beta_i$, conditioning the features before any Transformer layers \citep{FiLM}. We then feed each detector’s modulated tensor through the same pretrained Whisper encoder (convolutional stem, sinusoidal positional encoding, multi-head attention) loaded via HuggingFace \citep{HuggingFace}, extract the final hidden-state vector from each, and concatenate them. That concatenated vector passes through an MLP classifier (512 → 256 → 128 → 64 → 2 units, with softmax) \citep{MLP} to produce detection logits. \\

Since Whisper's encoder was originally trained on log‑mel spectrograms of 30 s audio clips, we fine‑tuned its weights using parameter‑efficient fine‑tuning (PEFT) \citep{PEFT}, a popular family of techniques used for adapting large language models to specific downstream tasks. Particularly, we employ DoRA (Weight‑Decomposed Low‑Rank Adaptation) \citep{DoRA}, which applies low‑rank updates to the output projection matrices of the multi‑head attention layers, tuning only 0.5\% of the 39 million parameters in the Whisper‑tiny architecture \citep{Whisper}. This strategy retains the pretrained model’s strengths while substantially reducing computational cost. In addition, we performed self-supervised contrastive pre-training so that the Q-Adapter and Whisper encoders learn noise-invariant, signal-sensitive features from unlabeled data, thereby accelerating and strengthening the subsequent supervised fine-tuning step for GW detection. Details of the pre-training and fine-tuning processes, and the model architecture are provided in the Methods section.

\begin{figure}[t!]
    \centering
    \includegraphics[scale=0.25]{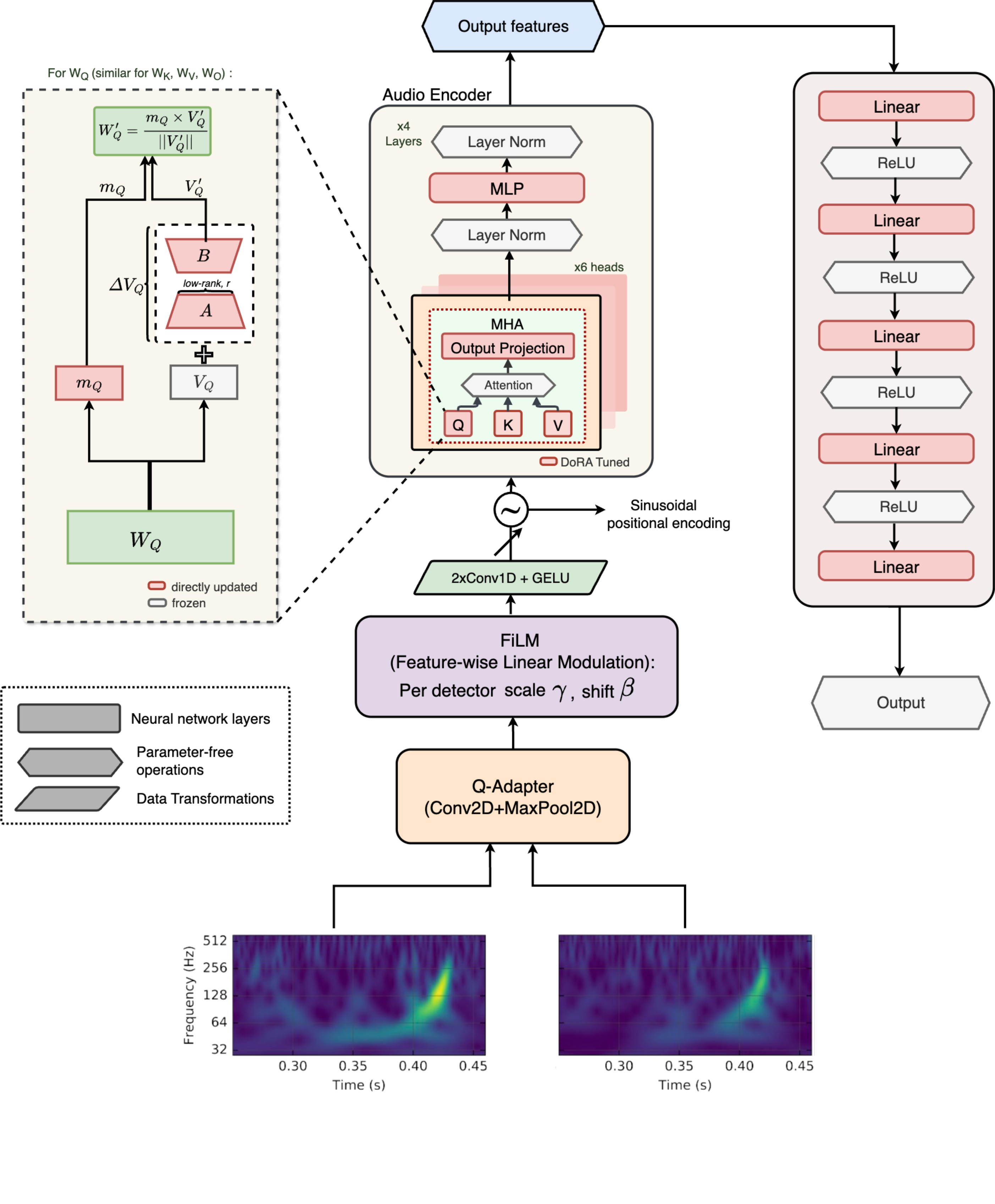}%
    \caption{Architecture of GW-Whisper. Raw strain from each detector is first converted into a Q‑transform spectrogram (via QScan) and fed through a lightweight convolutional ``Q‑Adapter,” module that extracts features from the QScans. Then feature‑wise linear modulation (FiLM) applies a learned per‑detector scale $\gamma $ and shift $\beta$ that learns detector-specific modulations. Sinusoidal positional encodings and 1-D convolutional layers project the pooled features for input to the frozen Whisper encoder network, whose multi‑head attention blocks and output‐projection weights are fine‑tuned with low‑rank DoRA adapters. The resulting per‑detector embeddings are concatenated and passed through a multilayer perceptron classification head to produce the final detection logits.}
    \label{fig:Whisper_architecture}
\end{figure}

\section{Results}\label{sec2}

We evaluate the performance of GW-Whisper on simulated data from the Machine Learning Gravitational-Wave Search Challenge (MLGWSC-1) \citep{ML_MDC}, which provides a benchmark for comparing GW search algorithms on standardized datasets. The challenge consists of four datasets with progressively increasing complexity; in this study, we focus on Dataset 3 and Dataset 4. Dataset 3 contains Gaussian noise with varying power spectral densities (PSDs) and injected GW signals that incorporate both precession effects and higher-order modes, using the IMRPhenomXPHM waveform family \citep{IMRPhenomXPHM}. Dataset 4 is the most realistic among the challenge sets, comprising real detector noise from the O3a observing run and GW signals simulated with precession and higher-order modes. Both datasets use two-detector inputs (from H1 and L1), and signals span up to 20 seconds in duration with component masses ranging from 7–50 M$_{\odot}$, spin magnitudes between 0 and 0.99, and isotropic spin orientations. Both of these test sets are 1 month in duration.\\

To quantify GW-Whisper’s detection performance, we compute the sensitive distance as a function of the False Alarm Rate (FAR) over the 1-month datasets. The sensitive distance, denoted \( D_{\rm sens}(F) \), is defined as the radius of a sphere with volume equivalent to the effective detection volume \( V(F) \) of the model at a given FAR \( F \). The detection volume is computed as:

\begin{equation}
V(F) \approx V(d_{\max}) \cdot \frac{1}{N_{\rm inj}} \sum_{i=1}^{N_{\rm found}(F)} \left( \frac{\mathcal{M}_{c, i}}{\mathcal{M}_{c, \max}} \right)^{5/2},
\end{equation}

where \( \mathcal{M}_{c, i} \) is the chirp mass of the \( i^{\text{th}} \) found injection, \( \mathcal{M}_{c, \max} \) is the upper limit of the injected chirp mass distribution, \( N_{\rm inj} \) is the total number of injections, and \( N_{\rm found}(F) \) is the number of injections recovered at the given FAR. The sensitive distance is then given by: \\

\begin{equation}
D_{\rm sens}(F) = \left( \frac{3 V(F)}{4\pi} \right)^{1/3}.
\label{eq:Sensitive distance}
\end{equation}

\begin{figure}[t!]
    \centering
    \includegraphics[scale=0.06]{Figure_2.pdf}%
    \caption{\label{fig:Sensitive_distance_vs_FAR} Sensitive distance versus FAR for MLGWSC‑1 dataset 3 (left) and dataset 4 (right). Solid curves correspon to ML-based algorithms, whereas the dotted lines represent traditional modelled (PyCBC) and unmodelled (cWB) searches. For dataset 4, results of an independent ML pipeline, SAGE, has been included. For both datasets, the most up-to-date results, based on latest publications from all search pipelines, wherever available, has been shown.}
\end{figure}

Following the evaluation protocol of MLGWSC-1, we process two separate datasets: one containing only noise (the background set) and another containing coincident signals and noise (the foreground set). Model outputs from both sets are post-processed using a clustering step, where individual segment-wise scores over the 1-month data are grouped into coherent triggers. Each trigger is then assigned a ranking statistic based on the maximum detection score in the cluster. We apply the Unbounded Softmax Replacement (USR) method of Schäfer et al. \citep{Training_strategies} to obtain a continuous ranking statistic from the network. During evaluation, each 2 h segment of the continuous 1-month data is whitened (via Welch’s method) \citep{Welch_method} and sliced into overlapping 1 s windows with a 0.1 s stride. The ranking statistics are obtained by evaluating the model on each slice with USR applied. Slices with ranking statistics exceeding a chosen threshold are flagged as first‑level triggers, which are then clustered into events by grouping any triggers within a small time window ($\Delta$t= 0.2 s). The times and values of the highest ranking statistic first-level trigger of each cluster is recorded from both the foreground and background data. The FAR at any ranking‑statistic threshold $R$ is computed by counting background events above $R$ and dividing by the 30 day live time. The detection efficiency at that threshold is determined by matching foreground events to injected signals within the clustering window. These quantities are combined, following Eq.~\ref{eq:Sensitive distance}, to compute the sensitive distance as a function of FAR. \\

Fig.~\ref{fig:Sensitive_distance_vs_FAR} (a) and (b) show our results on MLGWSC‑1 datasets 3 and 4, respectively, alongside the other pipelines from the challenge. Of the six submissions, four are machine‑learning–based: MFCNN, CNN‑Coinc, TPI FSU Jena \citep{TPI_FSU_Jena}, and Virgo‑AUTh (now AresGW) \citep{AresGW}. The other two are standard analyses: the matched‑filtering–based PyCBC \citep{PyCBC} and the unmodeled search cWB \citep{cWB}. For Dataset 4, we also include results from SAGE \citep{SAGE}, an independent machine learning (ML) algorithm that did not participate in the original challenge. Where available, we have incorporated the latest published results for these pipelines \citep{TPI_FSU_Jena, AresGW}. At a benchmark FAR of one per month, GW‑Whisper achieves a sensitive distance of $\sim 900$ Mpc on both datasets, outperforming the ML pipelines CNN‑Coinc and MFCNN, but remaining below cWB, PyCBC and the updated results for the other ML searches. For both datasets, GW-Whisper outperforms CNN-Coinc across all FARs, and achieves higher sensitivity than MFCNN up to FAR $\sim$ 200 per month for dataset 3, and $\sim$ 20 per month for dataset 4. Although GW‑Whisper delivers strong performance among ML‑based searches, its sensitivity is somewhat reduced by its reliance on fixed time–frequency spectrogram inputs, which is known to be sub-optimal for low SNR cases. Nevertheless, this approach remains particularly powerful for high‑mass BBH systems, whose merger and ringdown produce prominent, narrow‑band features that stand out clearly in the spectrogram domain. Looking forward, enhancing GW‑Whisper by incorporating multi‑resolution spectrograms or integrating raw time‑series convolution alongside the spectrogram encoder offer promising paths to improve sensitivity at low SNR while preserving the model’s computational efficiency and robustness.

\subsection{Application to O3b data}

\begin{figure*}[th!]
\centering
\includegraphics[scale=0.35]{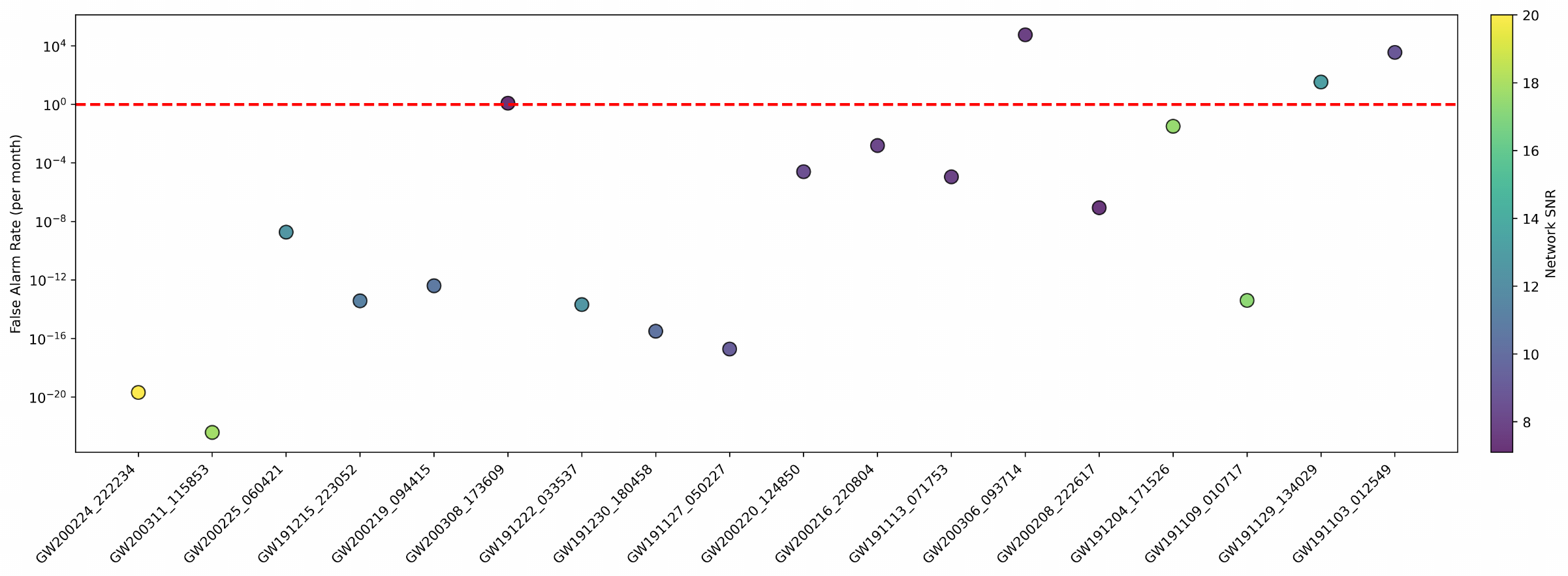}
\caption{\label{fig:Real_events} FAR per month for each BBH event in O3b, as recovered by GW-Whisper. The x-axis shows the event names, and each point is colored by their network SNR reported in \citep{GWTC-3}. The red dashed line marks FAR = 1 per month. Out of the 18 recovered events 16 lie below this threshold. }
\end{figure*}

We apply GW-Whisper on data from the second half of the third observing run (O3b) and cross match the events returned by our search against the BBH transients reported in GWTC-3 \citep{GWTC-3}. The O3b dataset was prepared exactly as outlined in the follow-up analysis by the TPI FSU Jena group \citep{TPI_FSU_Jena}: all segments of data flagged as science-quality and at least 60 s long were selected, with no exclusion of hardware injections, yielding 2377 independent segments totaling 8228706 s (95 days 6 hours) of coincident strain. We then cross-matched GW-Whisper’s candidate triggers against the reported events in GWTC-3 that fell within the O3b run, which were 31 confident and 4 marginal candidates. 
We mark a catalog event as found if the search returns an event within 0.2 s of reported time of the event in GWTC-3 catalog. The remaining catalog events are considered missed. Because we had only one month of O3a background to estimate the false-alarm distribution, many O3b candidates with very large ranking statistics lay in the zero-count regime where the empirical FAR is formally zero. To assign meaningful FARs below one per month, we adopted a hybrid extrapolation procedure.
For each candidate ranking statistic \(R\), we first compute the empirical FAR,

\begin{equation}
   \mathrm{FAR}_{\rm emp}(R) \;=\; \frac{N_{\rm bg}(R)}{T_{\rm bg}}, 
\end{equation}

where \(N_{\rm bg}(R)\) is the number of background clusters with statistic \(\ge R\) and \(T_{\rm bg}\) is the total background time (one month). If \(N_{\rm bg}(R)>0\), we set \(\mathrm{FAR}(R)=\mathrm{FAR}_{\rm emp}(R)\). Otherwise, we fit an exponential tail to the upper 5\% of the background distribution by performing a least-squares fit of $\ln \mathrm{FAR} \;=\; a\,x + b $
to the points \(\{(x_i,\,\mathrm{FAR}_{\rm bg}(x_i))\}\) in that tail, and then extrapolate

\begin{equation}
    \mathrm{FAR}(R) \;=\; \exp\bigl(a\,R + b\bigr)
\end{equation}

for zero-count candidates. This hybrid scheme yields a continuous FAR curve that matches the empirical background where available and provides sensible extrapolated FARs for the rarest O3b triggers. Fig.~\ref{fig:Real_events} shows the result of our search. We recover 18 events out of the 31 confident BBH events reported in GWTC-3, of which 16 events lies below FAR of one per month. The two events with FAR $>$ one per month both have network SNRs near 7–8 and are thus expected to be challenging. Meanwhile, some of the highest SNR events achieve as low FARs as $< 10^{-20}$ per month after extrapolation. \\

This result highlights the remarkable adaptability of foundation models, those originally trained on domains vastly different from GW astrophysics. Despite being pre-trained on log-mel spectrograms of human speech, the model, when adapted with domain-specific inputs (Q-scans), lightweight adapters, and PEFT, achieves strong performance on a highly specialized task like GW detection.

\subsection{Glitch classification}

In addition to signal detection, we evaluated GW-Whisper on the task of glitch classification and found that the vanilla Whisper configuration, using its original log-mel spectrogram front-end and without the QScan, Q-Adapter and FiLM layers, actually outperformed our GW-specialized variant. The standard log-mel inputs appear to preserve the broad, heterogeneous time–frequency patterns of transient noise more faithfully than the high-resolution QScans optimized for chirping signals. This suggests that for glitch identification, retaining the richer frequency coverage and learned filters of the original audio model can be more effective than our GW-specific front end. Similar to signal detection, we still implement PEFT with DoRA adapters and the same MLP architecture for glitch classification. \\


We fine-tuned GW-Whisper on single-detector simulated GW injections in O3 \citep{GWTC-3} noise, and 9 types of commonly-occurring glitches in the LIGO interferometers. These glitches were classified using the machine learning tool, GravitySpy \citep{GravitySpy, GravitySpy_classifier} with $>$ 90\% probability and reported SNR $\geq$ 8. The glitch data was obtained from the public data repository, Gravitational Wave Open Science Center (GWOSC) \citep{GWOSC}. Here, we preprocessed the whitened strain data by resampling the raw strains to 16 kHz and generating the (80 $\times$ 3000) dimensional log-mel spectrogram inputs for Whisper encoder. These spectrograms were generated using the off-the-shelf OpenAI WhisperFeatureExtractor module \citep{Whisper}. \\

\begin{figure}[t!]
\centering
    \includegraphics[scale=0.45]{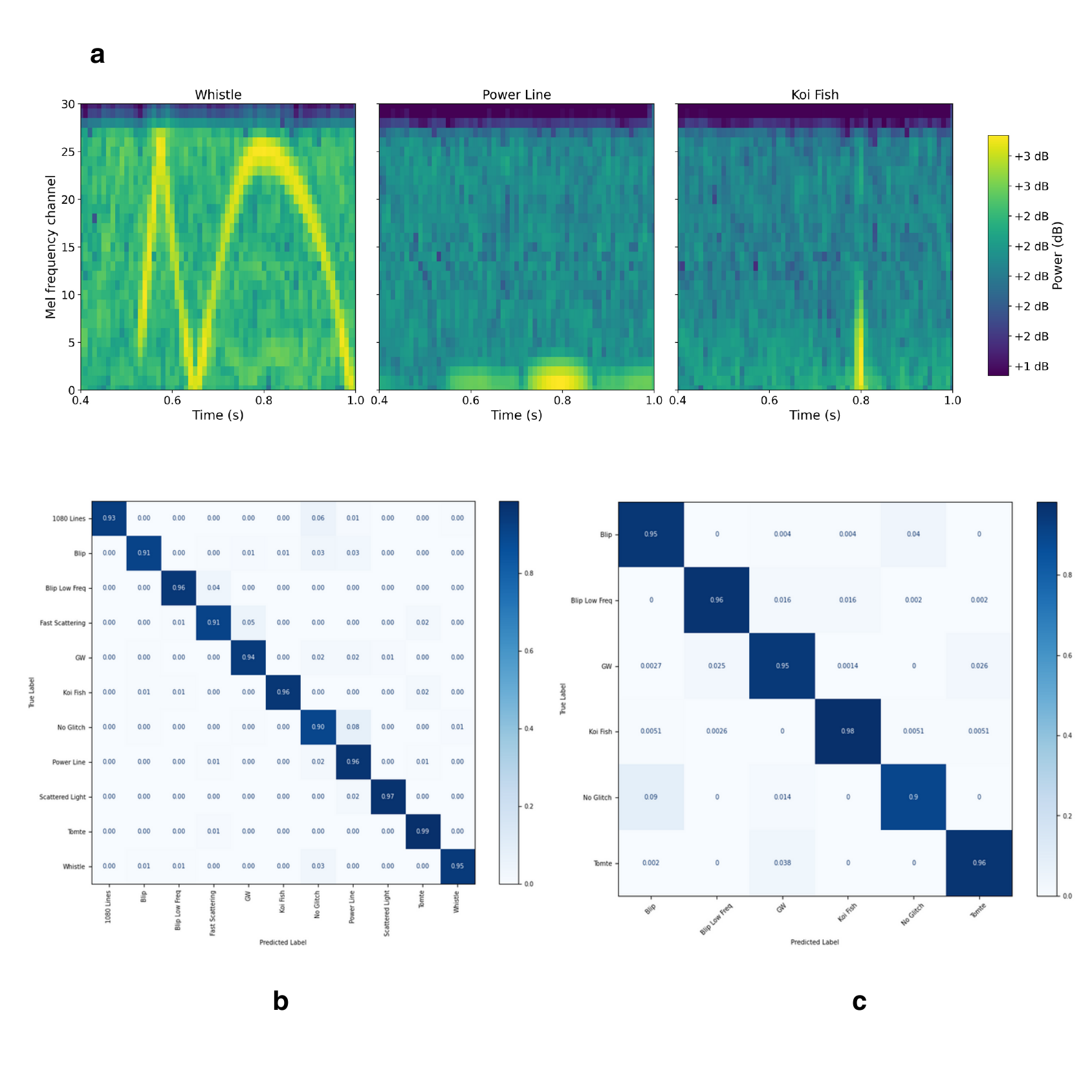}%
    \caption{\label{fig:Confusion_matrices} Top panel shows log-mel spectrograms of a Whistle, Power line and Koi fish glitch used in this study. Bottom panel shows confusion matrices describing the performance of GW-Whisper on the `generic' (left) and `high mass' (right) test cases. Each row of the matrices represents the true class, while each column represents the predicted class. The diagonal elements show the number of correct predictions and the off-diagonal elements indicate the misclassifications.}
\end{figure}

Fig.~\ref{fig:Confusion_matrices} (a) shows log-mel spectrograms of Whistle, Power line and Koi fish glitches. In order to evaluate how well GW-Whisper is able to distinguish various types of glitches from GWs, we conducted two different analyses. First we performed multi-label classification between vanilla BBH events (component masses between 10 and 50 M$_{\odot} $ and SNR between 6 and 15), and 9 types of glitches -- 1080 lines, Blip, Low-frequency blips, Fast scattering, Koi fish, Power line, Scattered light, Tomte and Whistle. An additional ``No glitch" class was included for samples where no prominent GW or glitch is visible. We call this analysis `generic' for the rest of the paper. In the second analysis, we considered classification between GWs with high total masses ($>$ 50 M$_{\odot}$) and glitches that often mimic such high mass events. Following the approach of \citep{GSpyNetTree}, we trained and tested GW-Whisper on simulated GW injections with total masses between 50 M${\odot}$ and 100 M${\odot}$, along with Blip, Low-frequency blip, Koi fish, and Tomte glitches. Throughout this paper, we refer to this analysis as the ``high-mass" study. This evaluation assesses GW-Whisper's ability to distinguish high-mass GW events from glitches with similar morphological features, as well as its capacity to differentiate between these signals.\\

The result of these tests is shown in Fig.~\ref{fig:Confusion_matrices} (b). The confusion matrices shown here evaluates GW-Whisper's multi-class classification performance for the `generic' case (left) and the `high-mass' case (right). A confusion matrix is a table that visualizes the performance of a classification model by comparing its predictions to the true labels. Each row of the matrix represents the true class, while each column represents the predicted class. The diagonal elements show the number of correct predictions for each class, while the off-diagonal elements indicate misclassifications between different classes. By analyzing the confusion matrix, one can identify which classes are being correctly classified, where the model struggles, and whether certain classes are frequently confused with others. The `generic' test shows that GW-Whisper achieves high classification accuracy across all classes, with true GW signals being correctly classified 94\% of the time. For the `high-mass' test, (Fig.~\ref{fig:Confusion_matrices} (c)), GW-Whisper achieves strong performance across all categories, with 95\% accuracy for correctly identifying GW signals, and near-perfect classification for Koi fish (98\%). Other glitches such as Blip, Low frequency blips, and Tomte are also classified with high accuracy, reaching 95\%, 96\% and 96\%, respectively. However, 9\% `no-glitch' samples are misclassified as blips, highlighting challenges in differentiating blip glitches from random fluctuations in pure noise samples. \\

These results show that GW-Whisper is a powerful tool to probe glitch families for characterizing and subtracting them from data. Having been trained on hundreds of thousands of hours of audio spanning myriad languages, accents, and acoustic environments, Whisper's early convolutional filters and attention layers have learned to extract salient features in the presence of diverse background noise. These learned invariances transfer naturally to transient glitches, which often resemble complex acoustic disturbances, allowing the model to distinguish subtle morphological differences without overfitting to any one glitch type. \\



\section{Discussion}\label{sec12}

This study demonstrates the application of GW-Whisper, an adaptation of OpenAI’s Whisper model, for GW data analysis, addressing critical challenges in GW astronomy. The novelty of this approach lies in its ability to adapt a speech recognition model for GW data without requiring full-scale retraining. By fine-tuning only a small fraction of the model’s parameters, GW-Whisper retains its pre-trained capabilities while efficiently adapting to the GW domain. This approach significantly reduces computational costs, compared to full finetuning of large foundation architectures, making it highly scalable for real-time data analysis in the era of increasing detection rates. By leveraging the pre-trained capabilities of Whisper, coupled with curated domain-specific modifications, GW-Whiaper's achieves robust signal detection on both simulated MLGWSC-1 benchmarks and real O3b observing-run data. \\

Intriguingly, for glitch classification, the original log-mel spectrogram front end (combined with DoRA and our MLP head) outperforms the GW-optimized Q-scan variant, underscoring that preserving the model’s audio-domain priors can better capture the heterogeneous structure of real detector glitches. At the same time, reliance on fixed spectrogram inputs imposes a modest sensitivity penalty at low network SNR, which explains why GW-Whisper trails matched-filtering (PyCBC) and the unmodeled search (cWB) in overall reach. Nevertheless, its minimal training cost and demonstrated robustness to evolving noise conditions make it an attractive complement to traditional pipelines—especially for high-mass systems ($\geq$ 100 $M_{\odot}$), where prominent merger–ringdown features dominate the spectrogram. This study further motivates the application of multi-modal foundation models in astrophysics, with potential extensions to other multi-messenger datasets. 



\section{Methods}\label{sec11}

\textbf{Data processing}:
We processed each raw strain segment by first whitening them by applying a Welch PSD estimate on 0.5 s blocks (with the inverse filter truncated to ±0.25 s), and then applying it to the full length of the data. During training and validation, BBH waveforms are generated with the IMRPhenomXPHM approximant \citep{IMRPhenomXPHM}, whitened using the same PSD, normalized to unit network signal-to-noise ratio, and then scaled to a random \(\rho_{\rm net}\in[7,20]\) before being added to noise windows. Pure-noise windows are labeled negative and noise + signal windows labeled positive. Table \ref{tab:injection-distros} summarizes the distributions of all intrinsic and extrinsic parameters used for injections.

\begin{table}[h]
\centering
\begin{tabular}{ll}
\toprule
\textbf{Parameter} & \textbf{Distribution} \\
\midrule
Waveform model & IMRPhenomXPHM \\
Component masses & \(m_1 \ge m_2 \sim \mathcal{U}[10,\,50]\,M_\odot\) \\
Spin magnitudes & \(\lvert \chi_{1,2}\rvert \sim \mathcal{U}[0,\,0.99]\) \\
Spin orientations & isotropic on the sphere \\
Coalescence phase \(\Phi_0\) & \(\mathcal{U}[0,\,2\pi)\) \\
Inclination \(\iota\) & \(\cos\iota \sim \mathcal{U}[-1,\,1]\) \\
Sky location \((\theta,\phi)\) & \(\sin\theta\sim\mathcal{U}[-1,\,1],\;\phi\sim\mathcal{U}[-\pi,\,\pi]\) \\
Polarization \(\Psi\) & \(\mathcal{U}[0,\,2\pi)\) \\
Sampling rate & \(2048\) Hz \\
Low-frequency cutoff & \(20\) Hz \\
Chirp distance \(d_c\) & \(d_c^2 \sim \mathcal{U}[130^2,\,350^2]\) Mpc\(^2\) \\
\bottomrule
\end{tabular}
\caption{Distributions of BBH injection parameters used during training and validation.  The chirp distance is defined by \(d_c = d\,(M_{c,0}/M_c)^{5/6}\) with \(M_{c,0}=1.22\,M_\odot\).}
\label{tab:injection-distros}
\end{table}

\textbf{Model Architecture}:
Whisper is an encoder-decoder architecture trained on multiple tasks to transcribe audio data into text \citep{Whisper}. The encoder portion of Whisper is a transformer-based architecture that embeds audio spectrograms into dense latent representations. It employs convolutional layers \citep{CNN_1, CNN_2} for initial feature extraction, sinusoidal positional encodings, and multi-head self-attention (MHA) layers within feedforward blocks \citep{Transformer}. This encoder processes audio signals into a representation that can be adapted to other downstream tasks by adding an additional model head. In this work, we make use of Whisper-tiny, the smallest in the suite of Whisper models. Further details about the size and architecture of this model can be found in Table 1 of \citep{Whisper}. The architecture of the Q-Adapter module described earlier is shown in Table \ref{tab:qadapter}.  \\

\begin{table}[h]
\centering
\begin{tabular}{lll}
\toprule
\textbf{Module} & \textbf{Input $\rightarrow$ Output} & \textbf{Description} \\
\midrule
QScan & $[B,T]\;\rightarrow\;[\,1\times F\times T\,]$ & Q-transform spectrogram (\texttt{QScan}) \\
Conv2d(1,32,3) & $[1\times F\times T]\;\rightarrow\;[32\times F\times T]$ & Frequency‐adapter conv layer \\
ReLU & $[32\times F\times T]\;\rightarrow\;[32\times F\times T]$ & Nonlinear activation \\
MaxPool2d(2) & $[32\times F\times T]\;\rightarrow\;[32\times \tfrac{F}{2}\times \tfrac{T}{2}]$ & Downsampling \\
Conv2d(32,64,3) & $[32\times \tfrac{F}{2}\times \tfrac{T}{2}]\;\rightarrow\;[64\times \tfrac{F}{2}\times \tfrac{T}{2}]$ & Conv layer \\
ReLU & $[64\times \tfrac{F}{2}\times \tfrac{T}{2}]\;\rightarrow\;[64\times \tfrac{F}{2}\times \tfrac{T}{2}]$ & Activation \\
MaxPool2d(2) & $[64\times \tfrac{F}{2}\times \tfrac{T}{2}]\;\rightarrow\;[64\times \tfrac{F}{4}\times \tfrac{T}{4}]$ & Downsampling \\
Conv2d(64,128,3) & $[64\times \tfrac{F}{4}\times \tfrac{T}{4}]\;\rightarrow\;[128\times \tfrac{F}{4}\times \tfrac{T}{4}]$ & Conv layer \\
ReLU & $[128\times \tfrac{F}{4}\times \tfrac{T}{4}]\;\rightarrow\;[128\times \tfrac{F}{4}\times \tfrac{T}{4}]$ & Activation \\
Conv2d(128,1,1) & $[128\times \tfrac{F}{4}\times \tfrac{T}{4}]\;\rightarrow\;[1\times \tfrac{F}{4}\times \tfrac{T}{4}]$ & Project back to 1 channel \\
AdaptiveAvgPool2d\,(80,3000) & $[1\times \tfrac{F}{4}\times \tfrac{T}{4}]\;\rightarrow\;[1\times 80\times 3000]$ & Resize to Whisper input shape \\
Scale \& Bias & $[1\times 80\times 3000]\;\rightarrow\;[1\times 80\times 3000]$ & Learned affine transform \\
FiLM ($\gamma,\beta$) & $[D\times 80\times 3000]\;\rightarrow\;[D\times 80\times 3000]$ & Detector-wise feature modulation \\
\bottomrule
\end{tabular}
\caption{Architecture of the Q-Adapter module.  Input batch size is $B$, $T$ is the time-series length, $F$ is the Q-transform frequency axis, and $D$ is the number of detectors.}
\label{tab:qadapter}
\end{table}

The QScans were generated on-the-fly using the GPU-accelerated implementation provided by the \texttt{ml4gw} library \citep{ml4gw}. In our Q-Adapter, we employ FiLM parameters to introduce detector-specific adjustments to the shared feature maps. After extracting and pooling the Q-transform output into a common tensor
\[
X \in \mathbb{R}^{D \times C \times T},
\]
where \(D\) is the number of detectors and \(C \times T = 80 \times 3000\) is the Whisper input shape, we apply a learned per-detector scale $\gamma \in \mathbb{R}^{D \times C}$ and shift $\beta \in \mathbb{R}^{D \times C}$. This allows each detector’s features to be dynamically rescaled and biased before being fed into the shared Whisper encoder, enabling the model to account for station-dependent sensitivity and noise characteristics. Specifically, if \(X \in \mathbb{R}^{C \times T}\) be the input feature map with \(C\) channels and \(T\) time–frequency bins, and \(\gamma \in \mathbb{R}^{C}\) and \(\beta \in \mathbb{R}^{C}\) are the learned scale and shift parameters, FiLM produces the modulated output \(\widetilde X\) by applying a learned scale \(\gamma \in \mathbb{R}^C\) and shift \(\beta \in \mathbb{R}^C\) per channel:

\begin{equation}\label{eq:film_per_element}
\widetilde X_{c,t} = \gamma_{c}\,X_{c,t} + \beta_{c},
\quad
c = 1,\dots,C,\; t = 1,\dots,T
\end{equation}

\begin{equation}\label{eq:film_vectorized}
\widetilde X = \gamma \odot X + \beta
\end{equation}

We apply DoRA \citep{DoRA} to the MHA blocks \citep{Transformer} of the original Whisper encoder architecture \citep{Whisper}. By fine-tuning only these trainable matrices while freezing the majority of the model's parameters, DoRA reduces memory and computational overhead while retaining the model's pretrained capabilities \citep{DoRA}. In our implementation of DoRA, we use rank-8 matrices to adapt the Query, Key, Value, and Output matrices of each MHA block. In doing so, we only modify a minuscule fraction of Whisper-tiny's parameters. Specifically, while the original model contains 39 million parameters, our DoRA implementation requires training just 196,608 parameters - approximately 0.5\% of the total model. \\

We append a custom classification head to the encoder output consisting of a fully connected network with three hidden layers having 1024, 512, 256 neurons respectively and ReLU activation between each layer, followed by a softmax to convert the outputs into probabilities. For signal vs. noise classification, the loss function used was binary cross-entropy, 

\begin{equation}
    \mathcal{L}_{\mathrm{BCE}}(y,\,\hat{y})
= -\bigl[y\,\log(\hat{y}) \;+\;(1 - y)\,\log\bigl(1 - \hat{y}\bigr)\bigr],
\end{equation}

where \(y\in\{0,1\}\) is the true label and \(\hat{y}\in(0,1)\) is the predicted probability, while for multiclass classification of GW and glitches, the categorical cross entropy was used:

\begin{equation}
    \mathcal{L}_{\mathrm{CCE}}(\mathbf{y},\,\hat{\mathbf{y}})
= -\sum_{k=1}^{K} y_{k}\,\log\bigl(\hat{y}_{k}\bigr),
\end{equation}
where \(\mathbf{y}=(y_{1},\dots,y_{K})\) is the one-hot true label vector and \(\hat{\mathbf{y}}=(\hat{y}_{1},\dots,\hat{y}_{K})\) are the predicted class probabilities. \\

\textbf{Contrastive Pre-training}:  To give GW-Whisper a head start on signal detection amidst diverse detector noise, we introduce a self-supervised contrastive pre-training step. We form mini-batches of $B$ positive pairs (two noisy renditions of the same injected waveform) and negative pairs (two pure-noise clips), and pass each view through our Q-Adapter and Whisper encoder to obtain embeddings
\[
h_1^{(i)},\,h_2^{(i)} \;\in\; \mathbb{R}^d,
\quad i=1,\dots,B.
\]
These are then projected via a small head $g:\mathbb{R}^d\to\mathbb{R}^p$:
\[
z_1^{(i)} = g\!\bigl(h_1^{(i)}\bigr),
\quad
z_2^{(i)} = g\!\bigl(h_2^{(i)}\bigr).
\]
After $\ell_2$-normalization, we form the $2B\times2B$ cosine-similarity matrix
\[
s_{ij} = \frac{z_i \!\cdot\! z_j}{\tau},
\]
where $\tau$ is a temperature hyperparameter.  The loss \citep{InfoNCE} is then
\[
\mathcal{L}_{\rm NCE}
= -\frac{1}{B}\sum_{i=1}^B
\biggl[
\ln\frac{\exp(s_{i,i+B})}{\sum_{k\neq i}\exp(s_{i,k})}
\;+\;
\ln\frac{\exp(s_{i+B,i})}{\sum_{k\neq i+B}\exp(s_{i+B,k})}
\biggr].
\]
Minimizing $\mathcal{L}_{\rm NCE}$ brings each positive pair $(z_1^{(i)},z_2^{(i)})$ closer in representation space while pushing all other (noise or different-injection) examples apart. Pre-training both the Q-Adapter and Whisper encoder with this contrastive objective teaches the network features that are invariant to noise realizations yet sensitive to the underlying waveform morphology.  When we subsequently fine-tune with DoRA adapters and our MLP classification head, these pretrained representations accelerate convergence, enhance robustness to non-stationary glitches, and reduce the need for large labeled training sets. \\

\textbf{Data availability:} The MLGWSC-1 datasets were generated using code from the following repository \citep{mlgwsc-1}. Data for the search on O3b data was generated using code from \citep{TPI_FSU_Jena_code}. The glitch data was obtained from open public data available in GWOSC \citep{GWOSC}. \\

\textbf{Code availability:} The code used for this analysis can be obtained from \citep{GW-Whisper_code}.




\backmatter





\bmhead{Acknowledgements}

This research was undertaken with the support of compute grant and resources, particularly the DGX A100 AI Computing Server, offered by the Vanderbilt Data Science Institute (DSI) located at Vanderbilt University, USA. This research used data obtained from the Gravitational Wave Open Science Center (https://www.gw-openscience.org), a service of LIGO Laboratory, the LIGO Scientific Collaboration and the Virgo Collaboration. LIGO is funded by the U.S. National Science Foundation. Virgo is funded by the French Centre National de Recherche Scientifique (CNRS), the Italian Istituto Nazionale della Fisica Nucleare (INFN) and the Dutch Nikhef, with contributions by Polish and Hungarian institutes. This material is based upon work supported by NSF's LIGO Laboratory which is a major facility fully funded by the National Science Foundation.


\bibliography{sn-bibliography}

\end{document}